\documentclass[prd,twocolumn,superscriptaddress,preprintnumbers,
noshowpacs,groupedaddress]{revtex4} %floatfix
\usepackage{epsfig}
\usepackage{color}
\usepackage{amsmath}
\usepackage{amssymb}
\usepackage{natbib}
\usepackage{enumerate}

\usepackage{graphicx}
\graphicspath{{figs/}}
\usepackage[caption=false]{subfig}
\usepackage{slashed}

\newcommand{\be}{\begin{equation}}
\newcommand{\ee}{\end{equation}}
\newcommand{\beq}{\begin{equation}}
\newcommand{\eeq}{\end{equation}}
\newcommand{\bea}{\begin{eqnarray}}
\newcommand{\eea}{\end{eqnarray}}
\newcommand{\beaa}{\begin{eqnarray*}}
\newcommand{\eeaa}{\end{eqnarray*}}
\newcommand{\ba}{\begin{array}}
\newcommand{\ea}{\end{array}}
\newcommand{\bi}{\begin{itemize}}
\newcommand{\ei}{\end{itemize}}
\newcommand{\ben}{\begin{enumerate}}
\newcommand{\een}{\end{enumerate}}

\begin{document}
%%%%%%%%%%%%%%%%%%%%%%%%%%%%%%%%%%%%%%%%%%%%%%%%%%%%%%%%%%%%%%%%%%%%%%
\preprint{ULB-TH/16-08}
%\title{New leptogenesis mechanism in the standard seesaw model at low scale}
\title{Higgs doublet decay as the origin of the baryon asymmetry}

\author{Thomas Hambye}
\email{thambye@ulb.ac.be}
\author{Daniele Teresi}
\email{daniele.teresi@ulb.ac.be}

\affiliation{Service de Physique Th\'eorique - Universit\'e Libre de Bruxelles, Boulevard du Triomphe, CP225, 1050 Brussels, Belgium}

%\date{\today}

%%%%%%%%%%%%%%%%%%%%%%%%%%%%%%%%%%%%%%%%%%%%%%%%%%%%%%%%%%%%%%%%%%%%%%
\begin{abstract}
We consider a question which curiously had not been properly considered so far: in the standard seesaw model what is the minimum value the mass of a right-handed (RH) neutrino must have for allowing successful leptogenesis via CP-violating decays?
To answer this question requires to take into account a number of thermal effects. We show that,  for low RH neutrino masses and thanks to these effects, leptogenesis turns out to proceed efficiently from the decay 
of the Standard Model (SM) scalar doublet components into a RH neutrino and a lepton. 
Such decays produce the asymmetry at low temperatures, slightly before sphaleron decoupling.
%We show that this is possible thanks to thermally induced one-loop CP-violating effects.
%Along this new mechanism, which proceeds slightly before sphaleron decoupling, 
If the RH neutrino has thermalized prior from producing the asymmetry, this mechanism turns out to lead to the bound $m_N>2$~GeV.
If, instead, the RH neutrinos have not thermalized, leptogenesis from these decays is enhanced further and can be easily successful, even at lower scales.
This Higgs-decay leptogenesis new mechanism 
%is particularly economical in terms of assumptions to be made to render it successful. For instance, it 
works without
requiring an interplay of flavor effects and/or 
cancellations of large Yukawa couplings in the neutrino mass matrix. Last but not least, such a scenario turns out to be testable, from direct production of the RH neutrino(s).
\end{abstract}

\maketitle
%%%%%%%%%%%%%%%%%%%%%%%%%%%%%%%%%%%%%%%%%%%%%%%%%%%%%%%%%%%%%%%%%%%%%%

%%%%%%%%%%%%%%%%%%%%%%%%%%%%%%%%%%%%%%%%%%%%%%%%%%%%%%%%%%%%%%%%%%%%%%
The origin of the matter-antimatter asymmetry of the Universe is one of the major phenomena that the Standard Model (SM) of 
fundamental interactions cannot account for.
Leptogenesis~\cite{Fukugita:1986hr} constitutes the most motivated explanation we have today for it. It is based on nothing but the ``seesaw" particles and interactions that one generally assumes to explain another major enigma beyond the SM: the origin of neutrino masses.
%the baryon asymmetry of the Universe.
If the seesaw states, whose decays are at the origin of the creation of the baryon asymmetry, are heavy, with a mass from $10^8-10^9$~GeV to the GUT scale, it is absolutely straightforward to make it successful in natural agreement with neutrino constraints; this mass scale typically holds as a lower bound for a hierarchical mass spectrum of RH neutrinos~\cite{hierarchical_bound}. However, for such seesaw scales, it is very difficult to conceive possible ways of testing this mechanism.
If the seesaw states lie instead at a more testable lower scale, leptogenesis is less generically successful, but still it can easily work if one make extra assumptions,  such as by invoking a quasi-degenerate mass spectrum for at least 2 RH neutrinos (the so-called resonant leptogenesis mechanism~\cite{Pilaftsis:1997jf,Flanz:1994yx,Pilaftsis:2003gt,Dev:2014laa}), or a cancellation of large Yukawa
%as to have at least 2 of the seesaw states to have deeply quasi-degenerate masses, or as from cancellation of large Yukawa 
couplings in the neutrino mass matrix.
%Thus a straigthforward question is, down to which mass of the seesaw states can we go for successful leptogenesis?
Thus two straightforward interesting questions, both for model building issues, as well as for experimentally probing leptogenesis (or at least the seesaw origin of the neutrino masses), are: a) down to which mass of the seesaw states can we go for successful leptogenesis? and b) up to which extent do we need to make these extra assumptions at low scale?

In the ordinary leptogenesis framework the creation of the asymmetry proceeds from the Yukawa-induced decay of one (or more) RH neutrino(s) into SM scalar and lepton doublets
\begin{equation}
{\cal L}\owns -\frac{1}{2} m_{N_\alpha} \overline{N}_\alpha^c N_\alpha -Y_{N_{\alpha i}} \widetilde{H}^\dagger \bar{N}_\alpha L_i + h.c. \;,
\end{equation}
with $L_i=(\nu_i,\,l^-_i)^T$, $H=(H^+,\,H^0   )^T$ and $\widetilde{H}=i \tau_2 H^*$.
That from such a decay there exists a lower bound on the mass of the decaying RH neutrino is clear. Beside the fact that below $T_{sph}=131.7$~GeV \cite{D'Onofrio:2014kta} the sphalerons sharply decouple, if the mass of the RH neutrino is below the sum of the masses of the final-state particles, the decay simply does not proceed.
However,  as will be discussed at length below, in this case the SM-scalar doublet component decays into a RH neutrino and a lepton are still open. These decays turn out to be able to  produce efficiently an asymmetry, thanks to thermal effects.
%may still proceed from the CP-violating decay of the scalar-doublet components 
%(or the Higgs boson $h$ and longitudinal $W$ and $Z$ at temperatures below the electroweak symmetry breaking temperature) 
%This requiresby incorporating
%finite-temperature effects.

\section{Low scale CP-violation dynamics}

For any low-scale leptogenesis scenario one can distinguish at least 3 epochs depending on whether the temperature is below or above the electroweak symmetry breaking (EWSB)
critical temperature $T_c\simeq 160$~GeV and the sphaleron decoupling temperature $T_{sph}=131.7$~GeV.
It is a good approximation to assume that the decoupling of the sphalerons is instantaneous, i.e.~above $T_{sph}$ the sphalerons are deeply in thermal equilibrium whereas below they do not occur.
% \cite{}.
The effect of EWSB, instead, is more progressive. For the vacuum expectation value (VEV) of the SM scalar field we consider 
$v(T)^2 \simeq (1 - T^2/T_c^2) \, \theta(T_c - T) \, v^2$, with $v= 246 \, \text{GeV}$,
as a result of the SM crossover \cite{Carrington:1991hz}.
%,Kajantie:1995kf
%\red{For the interplay of Goldstone and gauge bosons, below $T_c$ we should consider
%the decays into right-handed neutrinos and leptons of the $h,W^\pm,Z$ physical states, with the gauge bosons having both transverse and longitudinal components. As each of these states has a different mass, this implies a series of successive decays. 
%However, for the creation of the baryon asymmetry this is relevant only for $T_{sph}<T<T_c$, where these 4 states have approximately the same (thermal contribution included) masses.
%%since all those 4 states have approximately the same mass (especially when incorporating the thermal contribution to these masses, above the sphaleron decoupling temperature when the baryon asymmetry is created), 
%}
Below $T_c$ we should consider
the decays into RH neutrinos and leptons of the $h,W^\pm,Z$ physical states, rather than of the Higgs doublet.
However, for the creation of the baryon asymmetry this is relevant only for $T_{sph}<T<T_c$, where these 4 states have approximately the same (thermal contribution included) masses.
Thus, for simplicity in the following we will consider instead the decay of the full scalar doublet with the 4 states having the mass of the Higgs boson, which at $T=0$ is $m_H=m_h=125.6$~GeV. 

For the various particles, in particular the electroweak lepton doublet $L$, 
%the third-generation quark doublet $Q_3$ and the gauge bosons $W$ and $B$, 
the masses are approximately given by $m_i^2 \equiv m_i(T)^2 \simeq M_i^2(v(T)) + c_i T^2$,
where $M_i^2(v)$ is the VEV-dependent zero-temperature mass. The coefficients $c_i$ can be found e.g.~in Ref.~\cite{Giudice:2003jh}.
% and the Higgs VEV is given by
%\begin{equation}
%v(T)^2 \simeq \bigg( 1 - \bigg(\frac{T}{T_c}\bigg)^2 \,\bigg) \, \theta(T_c - T) \;,
%\end{equation}
%with $T_c \simeq 160 \, \mathrm{GeV}$. 
%The constants $c_i$ in \eqref{eq:th_masses} are given by
%\begin{align}
%c_L \ &= \ \frac{1}{32} (3 g^2 + g'^2)\;, \qquad & c_{Q_3} \ &= \ \frac{g_3^2}{6} + \frac{3 g^2}{32} + \frac{g'^2}{288} + \frac{h_t^2}{16}\;, \notag\\
%c_W \ &= \ \frac{11}{12} \, g^2\;, & c_B \ &= \ \frac{11}{12} \, g'^2 \;.
%\end{align}
Note that, given the small values of the RH neutrino Yukawa couplings at low scale, the thermal corrections are negligible for the masses of the RH neutrinos, but not necessarily for their mass splitting when they are quasi-degenerate, see below.
%,zero temperature masses of the leptons are subdominant with respect to the thermal masses at temperatures above the sphaleron decoupling. 
%\section{Absolute bound}
%To study leptogenesis in the low-$m_N$ region, it is important to take into account the thermal masses of the SM particles.    
For the thermal mass of the Higgs doublet $m_H(T)$, instead, we will consider the result that is obtained from the second derivative of the thermal effective potential, as given e.g.~in Ref.~\cite{Espinosa:1993bs}.
%detailed in Appendix~\ref{app:higgs}. \textbf{DT: If there's not enough space, let's just give some references on the thermal potential we use}
%\col{DT: Between $T_c = 160$ GeV and $T_{sph} = 131.7$ GeV we currently neglect the electroweak SSB, i.e. that the Goldstone modes are not present anymore, and their role in the decays is taken by $W$ and $Z$. Is this OK? Or should we do something more sophisticated?}.

\begin{figure}
\includegraphics[width=0.3 \textwidth]{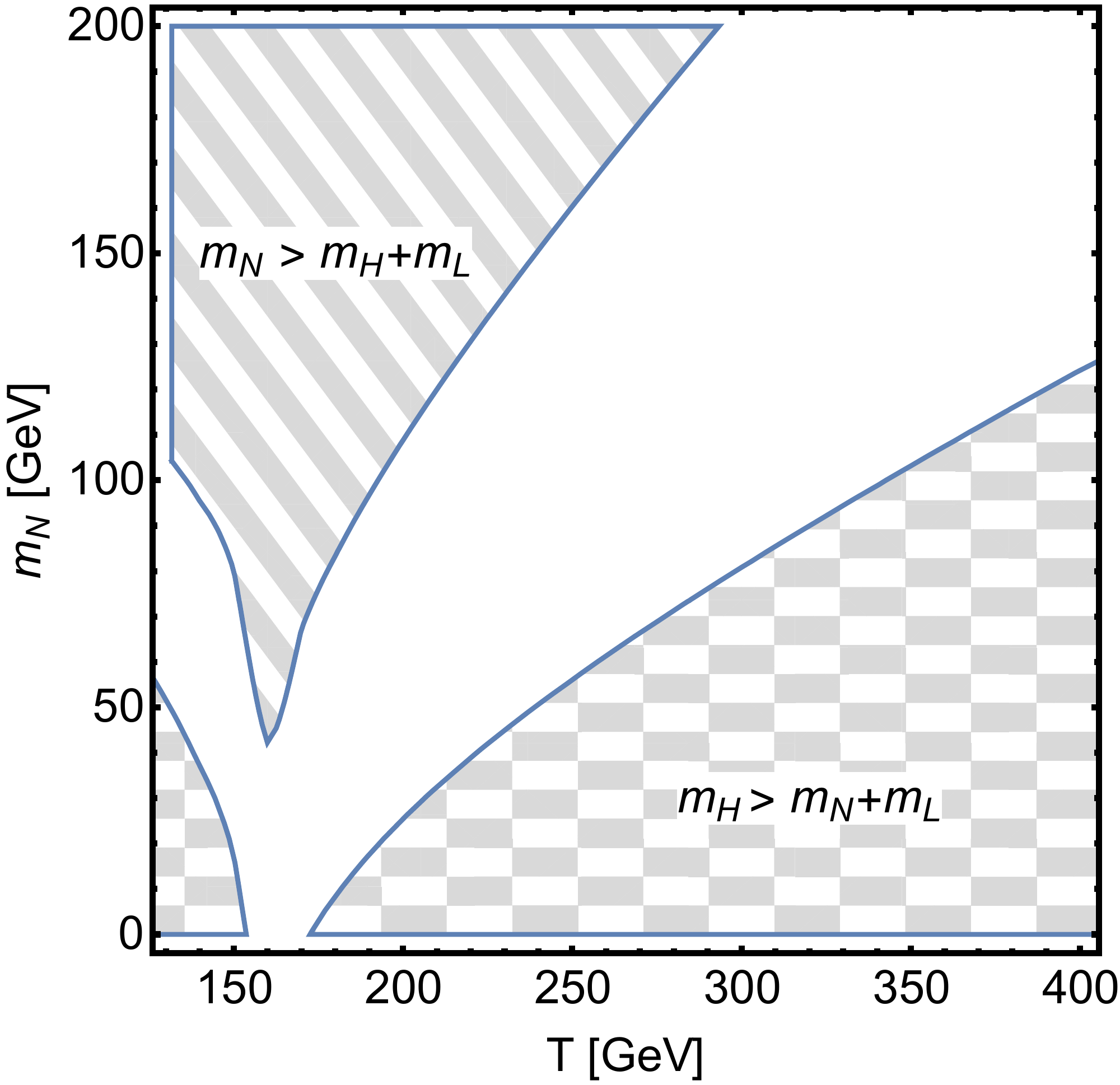}
\caption{Values of $m_N$ and $T$ for which the $N\rightarrow L H$ and $H\rightarrow L N$ decays are kinematically allowed. \label{fig:kinem_decays}}
\end{figure}

In Fig.~\ref{fig:kinem_decays} the regions in the $T$-$m_N$ plane in which the two different decay processes are active is shown.
For the moment, we work in the approximation of only one RH neutrino. Taking into account thermal masses, the decay widths for the processes $N \to L H$ and $H \to N L$ are respectively given by
\begin{align}
\Gamma_{N\to L H}  &=  \frac{m_N}{8\pi} \, Y_N Y_N^\dag \;   \lambda^{\frac{1}{2}}\!(1, a_H, a_L) \, (1- a_H + a_L) \;,\\
\Gamma_{H \to NL}  &=  \frac{m_N}{8 \pi} \, Y_N Y_N^\dag \; \lambda^{\frac{1}{2}}\!(1, a_H, a_L) \,  \frac{a_H - a_L - 1}{2 \, a_H^{3/2}} \;,
\end{align}
with $a_X \equiv (m_X(T)/m_N)^2$. We calculate the thermally-averaged decay rates $\gamma_{N\to L H}$ and $\gamma_{H \to NL}$ in the classical-statistics approximation, finding
\begin{align}
\gamma_{N\to L H}  &= \frac{m_N^3}{\pi^2 z} \, K_1(z) \, \Gamma_{N\to L H}   \;,\\
\gamma_{H\to N L}  &= \frac{m_H^2 \, m_N}{\pi^2 z} \, K_1\bigg(\frac{m_H}{m_N} z\bigg) \, 2\, \Gamma_{H\to N L} \;,
\end{align}
with $z \equiv m_N/T$. The total decay rate is thus given by $\gamma_D \ = \ \gamma_{N\to L H} \,\theta(m_N - m_H - m_L) \; + \; \gamma_{H\to NL} \,\theta(m_H - m_N - m_L)$. In the low-$m_N$ region the $\gamma_{H\to NL}$ rate receives $O(1)$ corrections~\cite{future} due to IR-enhanced processes involving electroweak bosons~\cite{IR}.

%\red{. The Boltzmann equations will be written in terms of the washout $\mathrm{K}$-factors, defined as 
%\begin{equation}
%\mathrm{K}_X \ \equiv \ \frac{\pi^2 z}{ m_N^3 K_1(z) \, H(z=1)} \, \gamma_X \;,
%\end{equation}
%with $z \equiv m_N/T$. We find
%\begin{align}
%\mathrm{K}_{N\to L H}  &= \frac{\widetilde{m}}{\widetilde{m}^*}\, \lambda^{\frac{1}{2}}(1, a_H, a_L) \, (1- a_H + a_L) \;,\\
%\mathrm{K}_{H\to N L}  &= \frac{\widetilde{m}}{\widetilde{m}^*}\, \lambda^{\frac{1}{2}}(1, a_H, a_L) \, \frac{a_H - a_L - 1}{a_H^{1/2}} \, \frac{K_1\big(\frac{m_H}{m_N} z\big)}{K_1(z)}\;,
%\end{align}
%where $\widetilde{m} = (Y_N Y_N^\dag)_{11}$ and $\widetilde{m}^* = 8 \pi 1.66 \,g_*^{1/2} \, v^2/m_{PL} \simeq 2.15\,  \mathrm{meV}$. The total decay factor is thus given by $\mathrm{K}_D \ = \ \mathrm{K}_{N\to L H} \,\theta(m_N - m_H - m_L) \; + \; \mathrm{K}_{H\to NL} \,\theta(m_H - m_N - m_L)$. In the low-$m_N$ region the $\mathrm{K}_{H\to NL}$ decay factor receives $O(1)$ corrections~\cite{future} due to IR-enhanced processes involving electroweak bosons~\cite{IR}.}

\begin{figure}
\includegraphics[width=0.18 \textwidth]{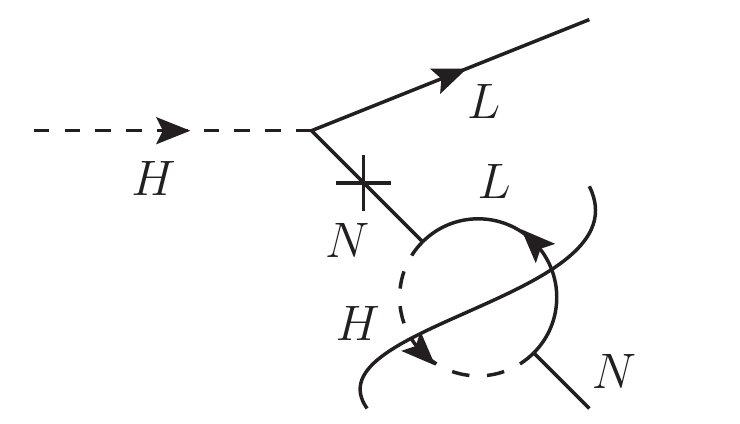}
\caption{Thermal cut in the $H \to NL$ decay, which gives rise to its purely-thermal L-violating CP-violation.\label{fig:graph}}
\end{figure}

The way the $H\rightarrow LN$ decays lead to a CP-asymmetry is from the one loop self-energy diagram of Fig.~\ref{fig:graph}. Clearly this diagram does not lead to any CP-violation at $T=0$, because the loop cannot have an absorptive contribution for $m_H> m_N+m_L$. However,
it does from thermal corrections, since one of the particles in the loop can originate on-shell from the thermal bath. Denominating by $\Gamma_T(z)$ the thermal cut of the self-energy, one obtains, for $|\Delta m_N(z)| \ll m_{N}$,  the usual resonant~\cite{Pilaftsis:1997jf,Flanz:1994yx}
%Covi:1996wh,Hambye:2003rt}
form for the unflavoured L-violating CP asymmetry \cite{Pilaftsis:1997jf,Frossard:2012pc} %(for $|\Delta m_N(z)| \ll m_{N}$)
\begin{equation}\label{eq:asym}
\epsilon_{CP}(z) \ = \ I_1 \;    \frac{2\,\Delta m_N^0 \Gamma_T(z)}{4\,\Delta m_N(z)^2 + \Gamma_T(z)^2} \;,
\end{equation}
 where $I_1 = Im[(Y_N Y_N^\dagger)^2_{12}]/(|Y_N Y_N^\dagger|_{11}|Y_NY_N^\dagger|_{22})$ and
$\Delta m_N(z) = \Delta m_N^0 + \Delta m_N^T(z)$ is the mass splitting including thermal corrections
\begin{align}\label{eq:therm_masses}
\Delta m_N^T(z) \ &\simeq \ \frac{\pi}{4 z^2} \, \Gamma_{22} \; \sqrt{\left(1 - \frac{\Gamma_{11}}{\Gamma_{22}}\right)^2 + 4 \,\frac{|\Gamma_{12}|^2}{\Gamma_{22}^2}} \nonumber\\
& \equiv \ \frac{\pi}{4 z^2} \, \Gamma_{22} \, f \;,
\end{align}
with $\Gamma_{ij}=m_N (Y_N Y_N^\dag)_{ij}/(8 \pi)$.
% to a good approximation, let us consider the mass splitting including thermal corrections $\Delta m_N(z) = \Delta m_N^0 + \Delta m_N^T(z)$, with
%\begin{align}\label{eq:therm_masses}
%\Delta m_N^T(z) \ &\simeq \ \frac{\pi}{2 z^2} \, \Gamma_{22} \; \frac{1}{2} \sqrt{\left(1 - \frac{\Gamma_{11}}{\Gamma_{22}}\right)^2 + 4 \,\frac{|\Gamma_{12}|^2}{\Gamma_{22}^2}} \nonumber\\
%& \equiv \ \frac{\pi}{2 z^2} \, \Gamma_{22} \, f \;,
%\end{align}
%and the resonant~\cite{Pilaftsis:1997jf,Flanz:1994yx,Covi:1996wh,Hambye:2003rt} maximal CP-asymmetry form~\cite{Pilaftsis:1997jf,Frossard:2012pc}
%\begin{equation}\label{eq:asym}
%\epsilon(z) \ = \ \frac{\Delta m_N^0 \Gamma_T(z)}{\Delta m_N(z)^2 + \Gamma_T(z)^2} \;,
%\end{equation}
%with $\Gamma_{ij}=m_N (Y_N Y_N^\dag)_{ij}/(8 \pi)$.
%Although a detailed analysis of the thermal CP-violation due to Higgs decays in the strongly resonant regime, 
Here we have conservatively taken the regulating expression in the denominator to be equal to the same $\Gamma_T(z)$ as in the numerator. This %guarantees that the asymmetry can take on its maximal value $1/2$, and 
is based on the physical expectation that the divergence in the degenerate limit is regulated by the (thermal) width of the heavy neutrinos. 
Notice that corrections to the precise form of the asymmetry (e.g. taking into account heavy-neutrino oscillations at $T\sim m_N$ in addition to mixing \cite{Dev:2014laa,Kartavtsev:2015vto}) 
can be absorbed into a re-definition of $f$.
As shown in detail in \cite{Hohenegger:2014cpa}, the masses appearing in the numerator of \eqref{eq:asym} should be taken as the Lagrangian masses without thermal corrections, $\Delta m_N^0$. This also guarantees the vanishing of the asymmetry in the CP-conserving limit $\Delta m_N^0 \to 0$.
The thermal cut of the Majorana RH neutrino self-energy has been calculated in \cite{Giudice:2003jh,Frossard:2012pc}.
%In \eqref{eq:asym} $\Gamma_T(z)$ is the thermal cut of the Majorana self-energy, calculated in \cite{Giudice:2003jh, Frossard:2012pc}. \
Here, neglecting the thermal motion of the decaying particle ,
we use the results
%this cut is purely thermal, since it is not kinematically allowed at $T=0$, Fig.~\ref{fig:graph}. 
%Here, we use the results\footnote{Having corrected for some typos in the formulas reported in Appendix D of~\cite{Frossard:2012pc}.} 
of \cite{Frossard:2012pc}, obtained in the Kadanoff-Baym formalism (which corresponds to taking the cut of the \emph{retarded} self-energy, rather than of the time-ordered one, as done instead in \cite{Giudice:2003jh}). 
%Note the factor $z^2$ difference with respect to \cite{Frossard:2012pc}}. 
The temperature dependence of $\Gamma_T$ can be extracted as $\Gamma_T(z) \equiv \Gamma_{22} \, \gamma(z)$, 
where $\gamma(z)$ is \cite{Frossard:2012pc}
\begin{equation}
\gamma(z) \ \equiv \ \frac{p L_\rho(q)}{p q} \;,
\end{equation}
with $p$ and $q$ the 4-momenta of the charged lepton and RH neutrino, respectively. The absorptive function $L_\rho(q)$ is given by
\begin{align}
L_\rho(q) \ &= \ 16 \pi \! \int \! d \Pi_H^q d \Pi_L^p \, (2 \pi)^4 \delta^4(l)\, \slashed{p} \,B \;,
\end{align}
where the momentum and statistical factor are $l=p-k-q$, $B = 1 + f_H - f_L$ for RH neutrino decay, and $l=q-k-p$, $B=f_H + f_L$ for  $H$ decay (with $f_X$ the corresponding Bose-Einstein or Fermi-Dirac distribution), and $d \Pi$ denoting the phase-space integration.  Thus, $L_\rho$ for both decays are as given in Appendix D of \cite{Frossard:2012pc} except that for $H$ decay we find that the $J_0$ term in $\vec{L}_\rho$ in  \cite{Frossard:2012pc} must be multiplied by $z^2$.
Thus the asymmetry \eqref{eq:asym} takes on the form
\begin{equation}\label{eq:asym2}
\epsilon_{CP}(z;x,f) \ = \ I_1 \; \frac{x \, \gamma(z)}{\left(x + \frac{\pi}{4 z^2} f\right)^2 \;+\; \gamma(z)^2} \;,
\end{equation}
where $x \equiv 2\,\frac{\Delta m_N^0}{\Gamma_{22}}$.

%In detail, we include $\Delta L = 1$ processes $LN \leftrightarrow Q_3 U_3$, $\bar{U}_3 N \leftrightarrow Q_3 \bar{L}$ , $\bar{Q}_3 N \leftrightarrow U_3 \bar{L}$ (neglecting the small difference between the thermal masses of $Q_3$ and $U_3$), $N \bar{L} \leftrightarrow \Phi A$, $L \Phi \leftrightarrow N A$, $\bar{L} A \leftrightarrow N \Phi$ (with $A= W, B$ and keeping thermal masses only when they regulate IR-enhanced contributions). The rates can be found in \cite{Giudice:2003jh}. The $s-$ and $t-$channel rates involving Higgs exchange will be denoted by $\gamma_{H s}$ and $\gamma_{H t}$, respectively. Instead, the gauge scattering rates by $\gamma_{A s}$ and $\gamma_{A t}$. 
%The corresponding washout factors are plotted in Fig. \ref{fig:rates}, for an illustrative set of values $m_N= 90 \, \mathrm{GeV}$, $\widetilde{m} = 0.1 \, \mathrm{eV}$.

%\begin{figure}
%\includegraphics[width=0.35 \textwidth]{rates}
%\caption{\label{fig:rates} Decay (solid line) and scattering (dashed lines) washout factors as functions of the temperature for $m_N= 90 \, \mathrm{GeV}$, $\widetilde{m} = 0.1 \, \mathrm{eV}$.} 
%\end{figure}

The Boltzmann equations for the RH neutrino and the lepton asymmetry, including the effect of the processes discussed above, are \cite{Pilaftsis:2003gt,Giudice:2003jh,Dev:2014laa}
\begin{align}\label{eq:boltzmann}
&\frac{n_\gamma H_N}{z} \, \frac{d \eta_N}{d z} = \bigg(1 \, - \, \frac{\eta^N}{\eta_N^{\rm eq}} \bigg)\Big[ \gamma_D + 2 (\gamma_{Hs}+\gamma_{As})  \nonumber\\
&\qquad \qquad \quad\; + 4 (\gamma_{Ht} + \gamma_{At})\Big] \;,\\
&\frac{n_\gamma H_N}{z} \, \frac{d \eta_L}{d z} = \gamma_D \bigg[\bigg(\frac{\eta^N}{\eta_N^{\rm eq}}-1 \bigg) \epsilon_{CP}(z) - \frac{2}{3} \eta_L \bigg] \nonumber\\
& \qquad\quad -\frac{4}{3} \eta_L \bigg[2 (\gamma_{Ht}+\gamma_{At}) + \frac{\eta^N}{\eta_N^{\rm eq}} (\gamma_{Hs} + \gamma_{As})\bigg] \;,\label{eq:boltzmann2}
\end{align}
where $\eta_a \equiv n_a/n_\gamma$ and $H_N$ is the Hubble rate at $T=m_N$.
%\red{\begin{align}\label{eq:boltzmann}
%&\frac{d \delta \eta_N}{d z} = \frac{K_1(z)}{K_2(z)} \, \bigg[1 + \delta \eta_N 
%- z \delta \eta_N \Big( \mathrm{K}_D \nonumber\\& \qquad\qquad + 2 (\mathrm{K}_{Hs}+\mathrm{K}_{As}) + 4 (\mathrm{K}_{Ht} + \mathrm{K}_{At})\Big)\bigg] \;,\\
%&\frac{d \eta_L}{d z} = \frac{z^3 K_1(z)}{2 \,\zeta(3)} \, \bigg[\mathrm{K}_D \Big(\delta \eta_N \epsilon_{CP}(z) - \frac{2}{3} \eta_L \Big) - \frac{4}{3} \eta_L \times \nonumber \\ &\Big(2 (\mathrm{K}_{Ht}+\mathrm{K}_{At}) + (1+\delta \eta_N) (\mathrm{K}_{Hs} + \mathrm{K}_{As})\Big)\bigg] \;,\label{eq:boltzmann2}
%\end{align}
%where $\eta_a \equiv n_a/n_\gamma$ and $\delta \eta_N \equiv \eta_N/\eta_N^{\rm eq} - 1$.}
These equations take into account additional important washout terms, which are active 
%It is also important to include washout due to scattering processes, which are active 
also when the decay processes are kinematically forbidden. For them we adopt the results and the notations of~\cite{Giudice:2003jh}, where they are calculated including the leading thermal effects. The final asymmetry produced in this way depends on 5 parameters: $m_N$, $I_1$, $x$, $f$  and the effective neutrino mass $\widetilde{m} \equiv v^2 (Y_N Y_N^\dag)_{11}/m_N$.

\section{Lower bound on $m_N$ for a thermalized $N$}

At first sight one could believe that the Boltzmann equations above do not lead to a lower bound on the mass of the RH neutrino, since the lower is $m_N$, the larger is the phase space available for the $H\rightarrow N L$ decay to occur. However, there exists one. Here, the out-of-equilibrium Sakharov condition is not realized as usual from the fact that the decaying particle is not in thermal equilibrium (here it is) but from the fact that the RH neutrino in the decay product is not. Thus, for $m_N< T_{sph}$, the lower $m_N$, the more $N$ is in thermal equilibrium at $T>T_{sph}$, the less successful is leptogenesis.

\begin{figure}
\includegraphics[width=0.35 \textwidth]{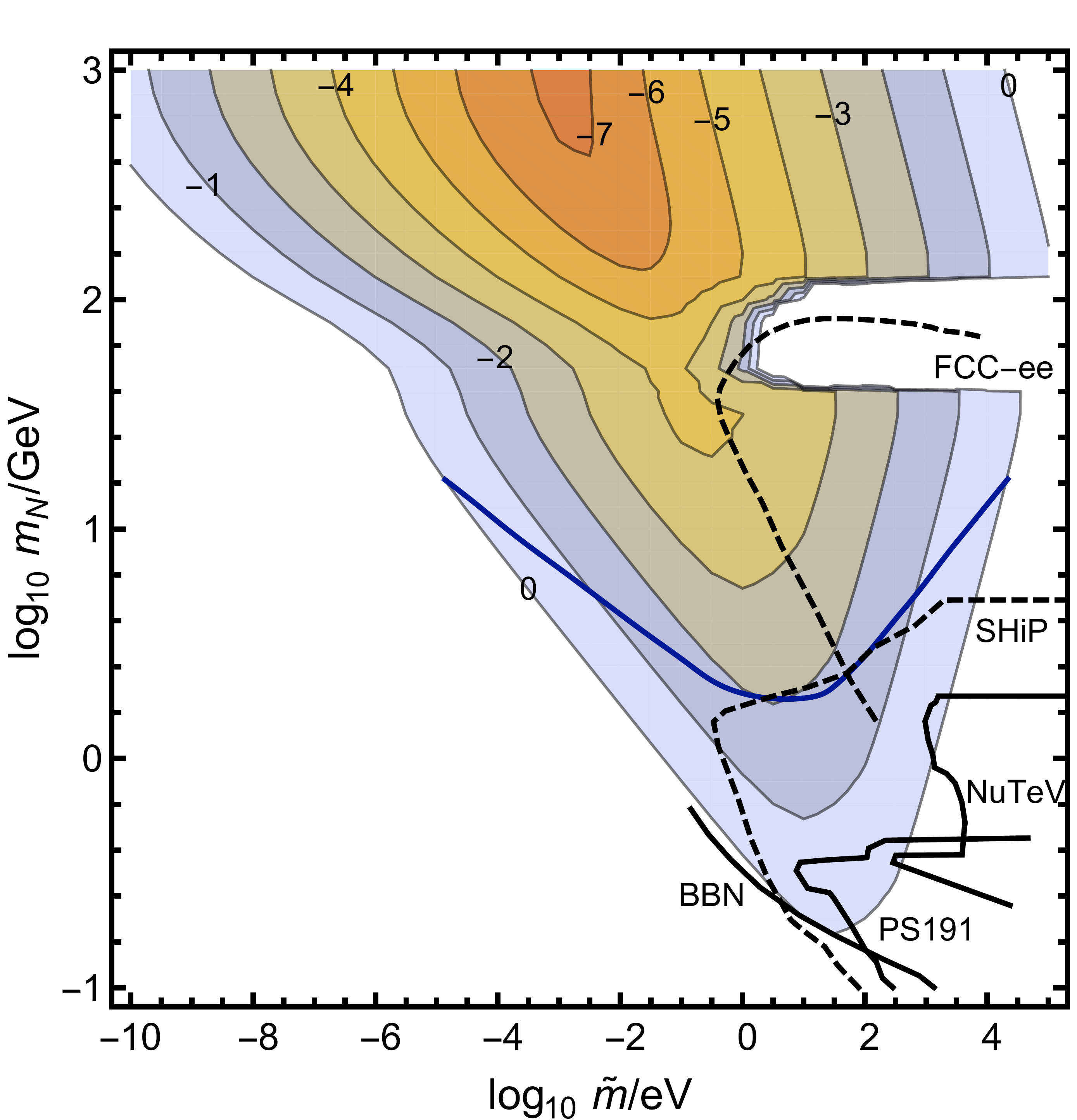}
\caption{Logarithm base 10 of the asymmetry $\epsilon_{CP}$ needed to obtain successful leptogenesis, with the RH neutrinos initially at thermal equilibrium. 
%For details, see text.
%{\bf $\epsilon$ is taken to be 0 when neither of the $N \to LH$ or $H \to NL$ decays are kinematically allowed.} 
We also plot the relevant existing bounds (solid lines) and projected sensitivities of the SHiP~\cite{Alekhin:2015byh} and FCC-ee~\cite{FCC} experiments (dashed lines). The area below the thick blue line requires values of $\epsilon_{CP}$ which are not reachable for such low $m_N$. \label{fig:bound}}
\end{figure}

Starting from a RH neutrino initially in equilibrium, Fig.~\ref{fig:bound} shows the results we get from solving
the Boltzmann equations~\eqref{eq:boltzmann}-\eqref{eq:boltzmann2}, by taking $\epsilon_{CP} = 10^{0,-1,-2,\ldots}$ when one of the two decay processes is kinematically allowed, zero otherwise. 
%In Fig.~\ref{fig:bound} we plot our results for the RH neutrinos initially at equilibrium. 
Taking the maximal CP-asymmetry $\epsilon_{CP}=1/2 \times 2$, (the factor of 2 is to take into account the fact that such a maximal CP asymmetry is obtained in the quasi-degenerate case together with a second RH neutrino), we obtain the bound $m_N>0.2\,\hbox{GeV}$.
%Taking the maximal CP-asymmetry $\epsilon=1/2 \times 2$,
%%, arising from the diagram of Fig.~\ref{fig:graph}, 
%when either of the two decay processes are kinematically allowed (the factor of 2 is to take into account the fact that such a maximal CP asymmetry is obtained in the quasi-degenerate case together with a second RH neutrino),
%%whose maximal CP asymmetry is also 1/2
%we plot our results in Fig.~\ref{fig:bound}, for the RH neutrinos initially at equilibrium.
%This gives the bound
%\begin{equation}
%m_N>0.2\,\hbox{GeV} \;.
%\label{absolutebound}
%\end{equation}

Of course one could wonder if this bound can be saturated, i.e.~if taking $\epsilon_{CP}=1/2$ can be justified. 
Although $\epsilon_{CP}(z) = 1/2$ cannot be satisfied at all temperatures, see~\eqref{eq:asym2},
%Fig.~\ref{fig:bound}
 since the bound occurs for $\widetilde{m}$ much larger than the usual thermal-equilibrium critical value $\widetilde{m}^* = 8 \pi 1.66 \,g_*^{1/2} \, v^2/m_{PL} \simeq 2.15\,  \mathrm{meV}$, the asymmetry in this case depends mostly on $\epsilon_{CP}$ at temperatures close to $T_{\rm sph}$. Thus, we find that taking $\epsilon_{CP} = const$ can be justified in a large portion of the parameter space in Fig.~\ref{fig:bound}. However, this is not fully the case for the low-$m_N$ region. The full asymmetry of Eq.~(\ref{eq:asym2}) (including in particular the $I_1$ factor) turns out to be maximized for $f\simeq1$.
For such values of $f$, since $\gamma(z \ll 1) \approx 50$, for $m_N < 10 \, \text{GeV}$ the thermal-mass contribution in the denominator of \eqref{eq:asym2} are important. Thus, for $T$ close to $T_{sph}$, the asymmetry is maximized for $x \sim \pi f T_{sph}^2/(4 m_N^2)$, which gives
\begin{equation}
\epsilon_{CP} \lesssim \ \frac{4}{\pi} \, \frac{50 \, m_N^2}{f\, T_{sph}^2} \;.
\end{equation}
This excludes the area below the thick blue line in Fig.~\ref{fig:bound}, yielding to a bound one order of magnitude stronger
\begin{equation}
m_N> 2\,\hbox{GeV} \;.
\label{absolutebound}
\end{equation}
This bound can be compared to the much larger one that we get by considering only $N\rightarrow LH$ decays, which turns out to be $m_N>50\,\text{GeV}$ (as one could approximately guess from Fig.~\ref{fig:kinem_decays}).
%Then,  if one considers not too large values of $f$, the thermal corrections to the mass splitting are subdominant. Thus, by taking $x \sim \gamma(m_N/T_{\rm sph})$ in Eq.~(\ref{eq:asym2}),  $\epsilon_{CP}$ is equal to $1/2$ in the region most relevant for the generation of the asymmetry and the bound of Fig.~\ref{fig:bound} can be saturated.
Note also that possible flavor effects, disregarded above, do not sizeably change this bound because in the low-$m_N$ region, where the bound occurs, even if $\widetilde{m} \gg \widetilde{m}^\star$, it turns out that there is no large washout effects diminishing the asymmetry produced, due to the sphaleron cut.

%%%%%%%%%%%%%%%%%%%%%%%%%%%%%%%%%%%%%
\section{The non-thermalized case: an efficient low-scale mechanism}

As explained above, the bound of Eq.~(\ref{absolutebound}) holds for leptogenesis induced by CP-violating H-decays if the RH neutrinos previously thermalize. The fact that it is in general difficult to achieve leptogenesis at such low scale, at the origin of this bound, is easy to understand: the lower the masses, the more the RH neutrinos were in thermal equilibrium at $T>T_{sph} \gg m_N$.
However, this is true only if one assumes that the N species has thermalized before the lepton asymmetry is produced.
If instead the RH neutrinos have not thermalized the situation drastically changes. This can be easily the case as long as 
%one considers the weak washout regime ($\tilde{m}<...$) and as soon as 
there were no other interactions below the reheating temperature (such as involving a $W_R$ for instance). For low $m_N$ the production of the asymmetry is cut off at $T_{sph}>m_{h,W,Z}>m_N$. Therefore the less $N$ thermalizes, the 
smaller is $n_N$, the fewer inverse $H$ decays occur (unlike $H$ decays which occur anyway), the larger is
$n_N^{eq}-n_N\sim n_N^{eq}$, the larger is the L-asymmetry produced. Note that this is different from what happens for large $m_N \ggg T_{sph}$, where considering a situation with no $N$ after reheating renders  leptogenesis more difficult~\cite{Giudice:2003jh}. 
In this case, in the weak washout regime, as the asymmetry is produced long before sphaleron decoupling, the more $N$ there are in the thermal bath, the more 
%there will be 
$N$ decays occur to produce the L asymmetry at $T\sim m_N$.

%This suggests the following new framework/mechanism based on 3 key issues.

%First, as explained above, this framework is based on the fact that for $m_N<m_H$ leptogenesis can proceed from $H\rightarrow NL$ decay (thus one could call it ``Higgs decay leptogenesis''). Second, it relies totally on the previously disregarded CP-violating low-scale mechanism of thermal corrections in this decay because, as emphasized above, without them there is simply no CP-asymmetry in the $H\rightarrow N L$ decay. Third, it assumes that the $N$ species has not thermalized before producing the L asymmetry. 

%This can be easily achieved as soon as 
%there were no other interactions (such as involving a $W_R$ for instance) below the reheating temperature. 
%Note that,
%for high masses $m_N \ggg T_{sph}$, considering a situation where there were no $N$ after reheating renders  leptogenesis more difficult~\cite{Giudice:2003jh}. 
%In particular, in the weak washout regime, the more $N$ there are in the thermal bath, the more 
%$N$ decays are present to produce the L asymmetry at $T\sim m_N$. Here, at low scale, 
%the situation is rather opposite, because for low $m_N$ the production of the asymmetry is cut off at $T_{sph}>m_{h,W,Z}>m_N$. The less $N$ thermalizes, the 
%smaller is $n_N$, the fewer inverse $H$ decays occur (unlike $H$ decays which occur anyway), the larger is
%$n_N^{eq}-n_N\sim n_N^{eq}$, the larger is the L-asymmetry produced.

\begin{figure}
\includegraphics[width=0.35 \textwidth]{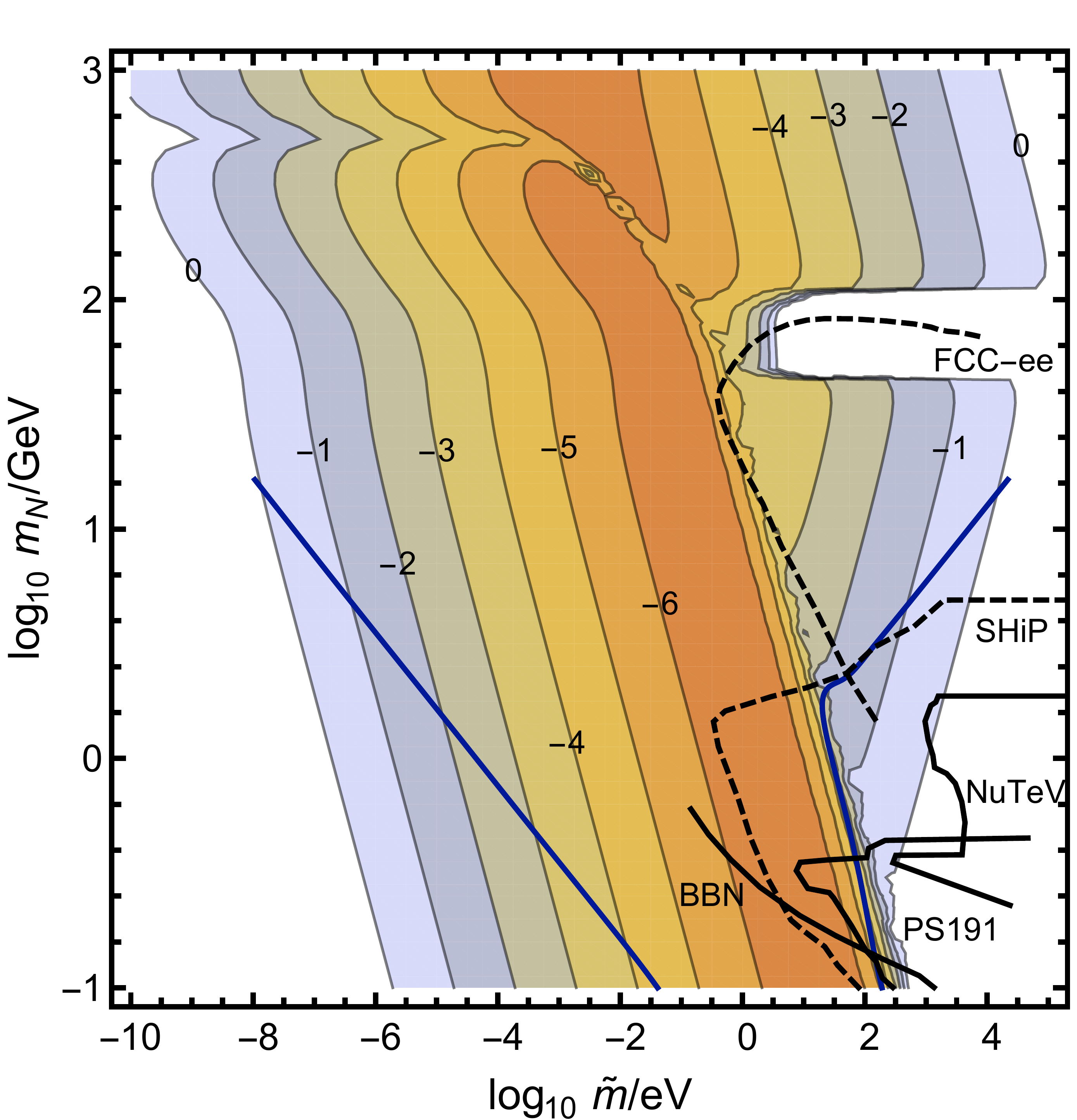}
\caption{Same as Fig.~\ref{fig:bound}, starting from no $N$ at $T_{in}=10\,T_{sph}$. \label{fig:empty}}
\end{figure}

Fig.~\ref{fig:empty} shows the numerical solution of the Boltzmann equations, by starting from a zero number density of RH neutrinos at $T_{\rm in} = 10 \, T_{\rm sph}$ and taking a maximal CP-asymmetry $\epsilon_{CP}=1/2$ (multiplied by 2 as above). Clearly this shows that, even for $m_N\sim 0.1$~GeV, the parameter space available is large and successful leptogenesis can be achieved with CP-asymmetry far from maximal. Note that here most of the asymmetry is created shortly before sphaleron decoupling because for $T \gg m_N$ and small $N$ number density, the source term in \eqref{eq:boltzmann2} is approximately constant: $d \eta/ d z \approx const$, since $\gamma_{D} \propto 1/z^4$ in this regime. Thus, the final asymmetry produced does not depend on the reheating temperature as long as this is larger than $T_{sph}$ by a factor of about 2. 
%Thus, the final asymmetry produced does not depend on the reheating temperature by more than a factor $\sim 2$ ($\sim 10$) as long as it is larger than $T_{sph}$ by a factor of about 2 (10). 

%Thus in the following we will focus on cases far from the resonant peak where the question of the regulator form is irrelevant. Thus ....
%\begin{equation}
%\epsilon_{CP}=...
%\end{equation} 

\begin{figure*}[t!]
\subfloat[]{\includegraphics[width=0.3 \textwidth]{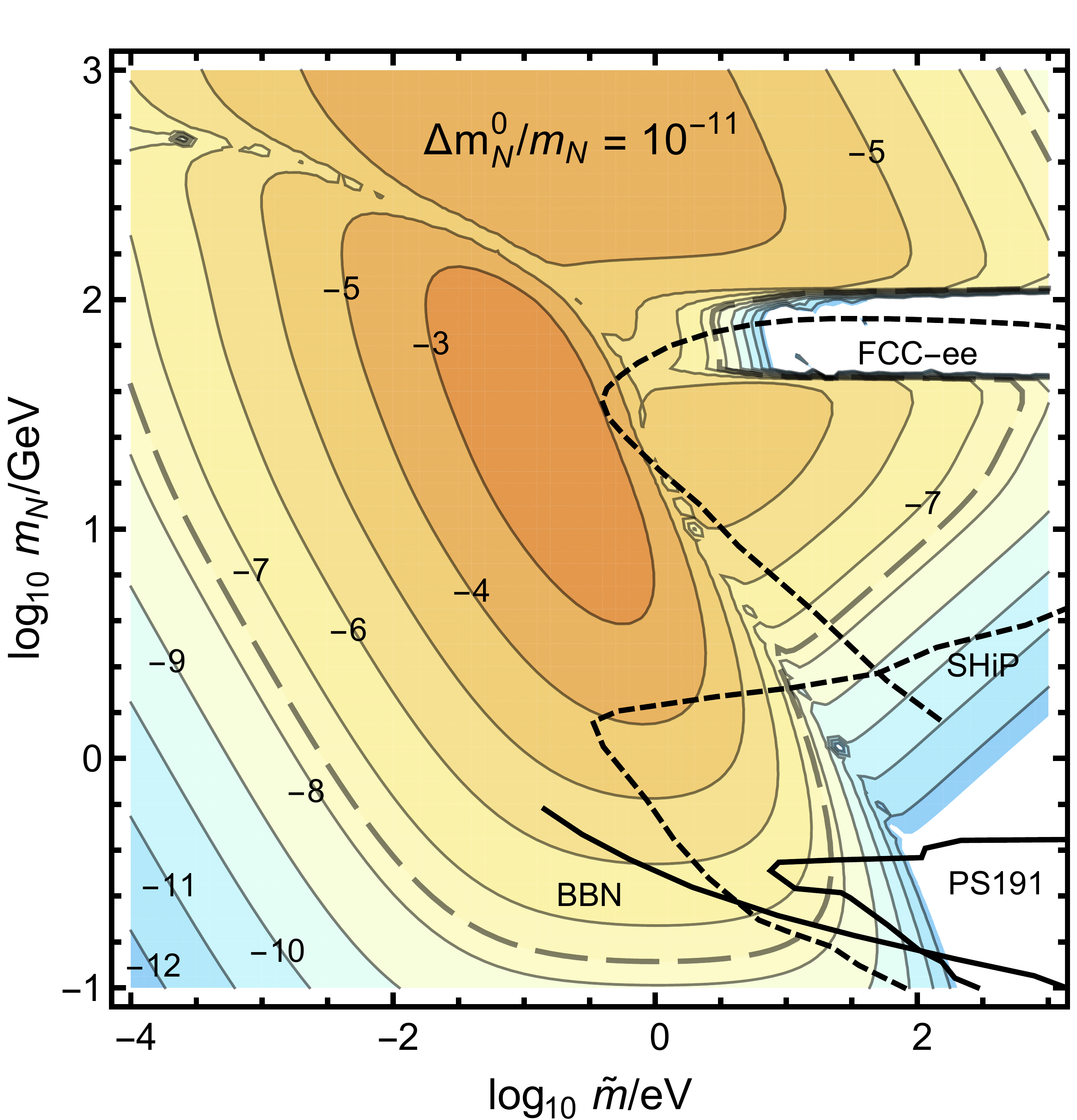}
\label{fig:dm11}}\quad
\subfloat[]{\includegraphics[width=0.3 \textwidth]{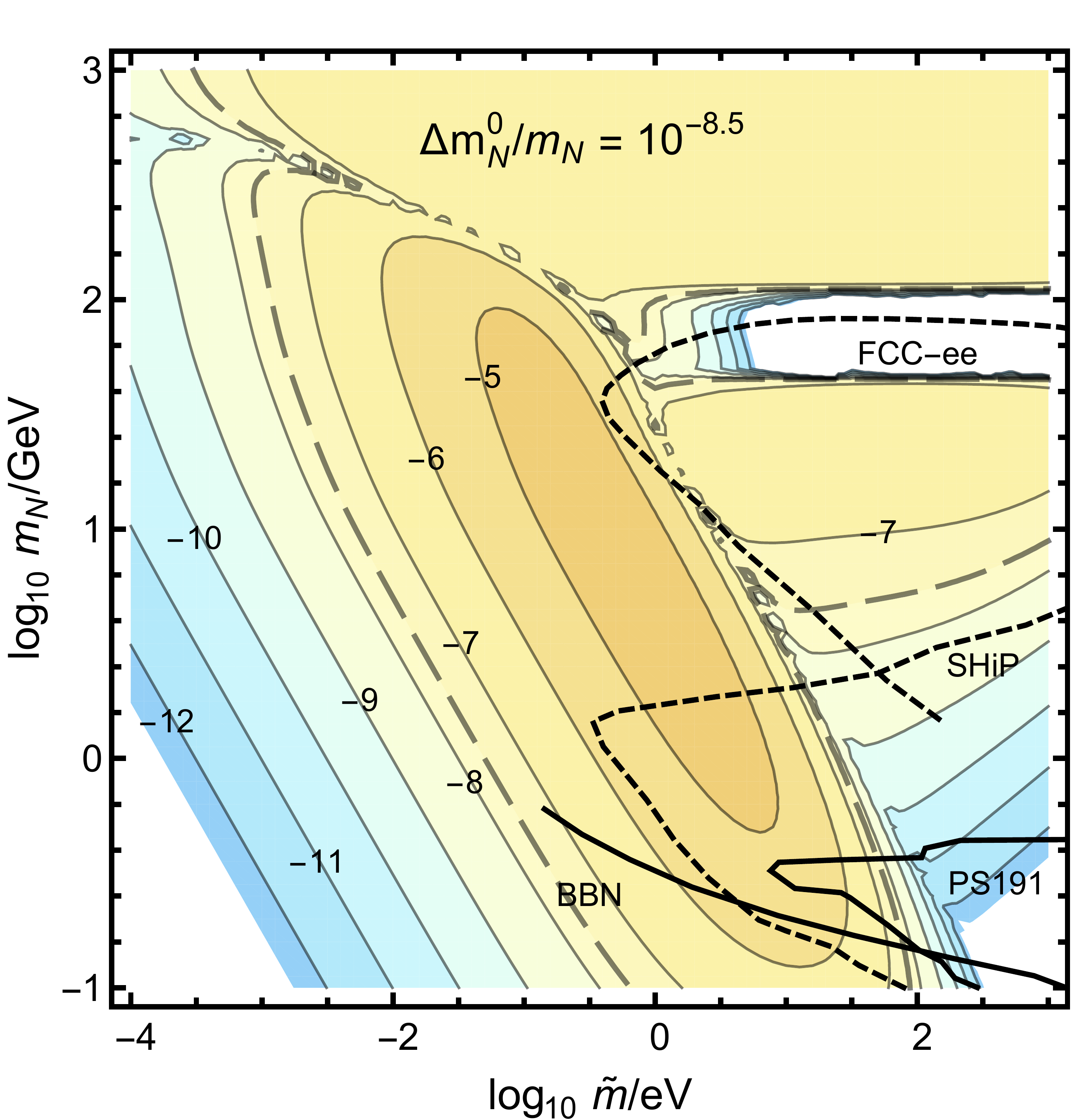}
\label{fig:dm8}}\quad
\subfloat[]{\includegraphics[width=0.3 \textwidth]{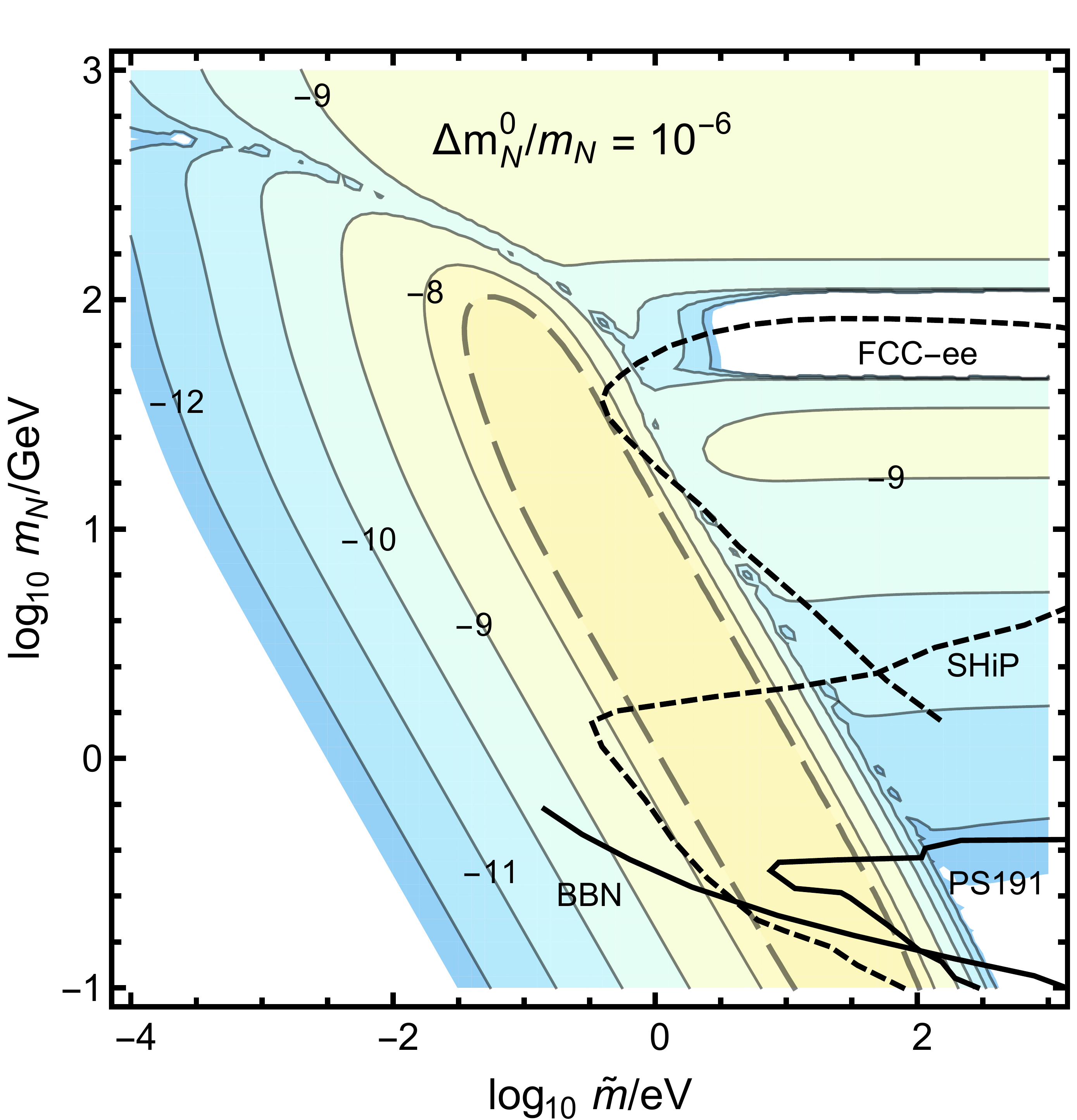}
\label{fig:dm5}}
\caption{Values of $\eta_L=n_L/n_\gamma$ obtained starting from no $N$ at $T_{\rm in} = 10 \, T_{\rm sph}$, for $\Delta m_N^0/m_N = 10^{-11,-8.5,-6}$. We have taken $\Gamma_{11}= (m_{\rm sol}/m_{\rm atm})\Gamma_{22}$ and $f=I_1=1$ for definiteness. The long-dashed line gives the minimum value needed $\eta_L^{\rm obs} = 2.47 \times 10^{-8}$.\label{fig:dm}}
\end{figure*}

In Figs.~\ref{fig:dm11}-\ref{fig:dm5} we plot the $\eta_L$ numerical solution of the Boltzmann equations, with the asymmetry as given in \eqref{eq:asym2}, with zero number density of RH neutrinos at $T_{\rm in} = 10 \, T_{\rm sph}$, and taking $\Delta m_N^0/m_N = 10^{-11,-8.5,-6}$ respectively.
%~\footnote{The unusual choice of the intermediate value has been maintained from Italian-cinema fanatic considerations.}
Fig.~\ref{fig:dm8} and~\ref{fig:dm5} show that successful leptogenesis, which requires $\eta_L > 2.47 \times 10^{-8}$, can be achieved with level of $N$ mass quasi-degeneracy about two orders of magnitude smaller than in ordinary TeV-scale resonant leptogenesis~\cite{Pilaftsis:2003gt,Dev:2014laa}. For $\Gamma_{11}/\Gamma_{22}=m_{\rm sol}/m_{\rm atm}$, we find that the minimum level of mass-degeneracy required is about $\Delta m_N^0/m_N \sim 10^{-5}$. In the flavoured (total-lepton-number conserving) mechanism considered in~\cite{Drewes:2012ma} with 3 RH neutrinos, which does not require $N$ mass degeneracy (and occurs at $T \sim 10^6 \, \text{GeV}$), a comparable level of fine-tuning is instead present in the Yukawa couplings to guarantee $\widetilde{m} \sim 10^3 \, \text{eV} \, \approx \, 10^5 \, m_{\rm sol}$, as required by the flavour effects taking place.

We may also compare the mechanism considered here with the ARS oscillation one~\cite{Akhmedov:1998qx} (which also relies on non-thermalized $N$, but with CP-violation given by N oscillations) in the $\rm \nu MSM$ scenario considered in \cite{Asaka:2005pn,Canetti:2012kh}. In this scenario, a mass degeneracy between 2 RH neutrinos of about $\Delta m_N^0/m_N \simeq 10^{-11}$ is needed to generate both the observed asymmetry (e.g.~at~$T \gg T_{sph}$) via the ARS mechanism and the dark matter relic density at $T \sim 100 \, \text{MeV}$, via $N$ freezeout or decay. 
The approximate form of the asymmetry  at $T_{ sph}$ generated by the ARS mechanism in this regime can be found in~\cite{Asaka:2005pn}.
%\begin{equation}
%\frac{\max \eta_L}{\eta_L^{obs}} \ \approx \ 0.02 \, \frac{(m_N/\text{GeV})^{5/3}}{(\Delta m_N^0/m_N)^{2/3}} \; \frac{\widetilde{m}_1 \widetilde{m}_2 (\widetilde{m}_1+\widetilde{m}_2)}{\text{eV}^3} \;.
%\end{equation}
Fig.~\ref{fig:dm11} shows that, assuming maximal CP-phases for both mechanisms, the asymmetry at $T_{sph}$ generated for $\Delta m_N^0/m_N \simeq 10^{-11}$ is about 7 (12) times larger than the ARS one, for $m_N = 2 \, (10) \, \text{GeV}$ (or larger if the reheating temperature is larger than $T_{sph}$ but smaller than the typical $T \gg T_{sph}$ ARS asymmetry production temperature).
Note that such a dominance of the asymmetry produced by H decays slightly before the sphalerons decouple does not hold for all the available parameter space \cite{future}. Notice also that, although the L-violating effects inducing the baryon asymmetry here can in principle be captured by the density-matrix formalism used to study the ARS mechanism (see e.g.~\cite{Canetti:2012kh} and~\cite{future}), these have been so far thought to be negligible and hence disregarded.

A remarkable feature of the framework considered in this letter is that, along it, leptogenesis is testable!
This is shown in all figures 3-5, which give the actual excluded $m_N$-$\widetilde{m}$ regions from various experiments, together with future expected sensitivities for $N$ production at SHiP and FCC-ee, see e.g.~\cite{Alekhin:2015byh, FCC}. Also shown is the lower bound on $m_N$ obtained from requiring that the decay of $N$ occurs before BBN. Clearly, for $m_N$ around GeV the available parameter space will be largely probed, with possibilities up to $\sim 50$~GeV. Note also that, as Fig.~\ref{fig:empty} suggests, if it was not for the BBN bound, leptogenesis from $H$ decays could be successful for values of $m_N$ smaller than considered here \cite{future}. For $f\sim1$ we find that leptogenesis can be successful for $m_N$ as low as $\sim20$~MeV.

To sum up, the leptogenesis from Higgs decay mechanism proposed here is particularly efficient at low scale, based on the 3 following ingredients. First it is based on the fact that, at low scale, thermal effects induce L-violating CP-violation in the decays of the SM scalar doublet into a RH neutrino and a lepton. Second it satisfies the out-of-equilibrium Sakharov condition from the fact that the RH neutrinos in the decay product (rather than the decaying particles) are out-of-equilibrium.
% First, it is based on the fact that for $m_N<m_H$ leptogenesis can proceed from $H\rightarrow NL$ decay (thus one could call it ``Higgs decay leptogenesis''). Second, it relies totally on the previously disregarded CP-violating low-scale mechanism of thermal corrections in this decay because, as emphasized above, without them there is simply no CP-asymmetry in the $H\rightarrow N L$ decay. 
Third, it assumes that the $N$ species has not thermalized before producing the L asymmetry, which boosts the asymmetry production. This mechanism is testable. 
%********************************************

%\vspace{-3mm}
%\begin{acknowledgements}

%\vspace{-2mm}

This work is supported by the FNRS-FRS, the FRIA, the IISN, an ULB-ARC and the Belgian Science Policy, IAP VI-11.


\begin{thebibliography}{99}


\bibitem{Fukugita:1986hr}
  M.~Fukugita and T.~Yanagida,
  %``Baryogenesis Without Grand Unification,''
  Phys.\ Lett.\ B {\bf 174} (1986) 45.

\bibitem{hierarchical_bound}
  S.~Davidson and A.~Ibarra,
  %``A Lower bound on the right-handed neutrino mass from leptogenesis,''
  Phys.\ Lett.\ B {\bf 535} (2002) 25
  [hep-ph/0202239];  T.~Hambye, Y.~Lin, A.~Notari, M.~Papucci and A.~Strumia,
  %``Constraints on neutrino masses from leptogenesis models,''
  Nucl.\ Phys.\ B {\bf 695} (2004) 169
  [hep-ph/0312203].
  %%CITATION = doi:10.1016/j.nuclphysb.2004.06.027;%%

\bibitem{Pilaftsis:1997jf}
  A.~Pilaftsis,
  %``CP violation and baryogenesis due to heavy Majorana neutrinos,''
  Phys.\ Rev.\ D {\bf 56} (1997) 5431
  [hep-ph/9707235].  

\bibitem{Flanz:1994yx}
  M.~Flanz, E.~A.~Paschos and U.~Sarkar,
  %``Baryogenesis from a lepton asymmetric universe,''
  Phys.\ Lett.\ B {\bf 345} (1995) 248
   Erratum: [Phys.\ Lett.\ B {\bf 382} (1996) 447]
  [hep-ph/9411366];
%\bibitem{Covi:1996wh}
  L.~Covi, E.~Roulet and F.~Vissani,
  %``CP violating decays in leptogenesis scenarios,''
  Phys.\ Lett.\ B {\bf 384} (1996) 169
  [hep-ph/9605319];
%\bibitem{Hambye:2003rt}

  
\bibitem{Pilaftsis:2003gt}
  A.~Pilaftsis and T.~E.~J.~Underwood,
  %``Resonant leptogenesis,''
  Nucl.\ Phys.\ B {\bf 692} (2004) 303
  [hep-ph/0309342]. 
%%\bibitem{Deppisch:2010fr}
%  F.~F.~Deppisch and A.~Pilaftsis,
%  %``Lepton Flavour Violation and theta(13) in Minimal Resonant Leptogenesis,''
%  Phys.\ Rev.\ D {\bf 83} (2011) 076007
%  [arXiv:1012.1834 [hep-ph]].  
  
\bibitem{Dev:2014laa}
  P.~S.~Bhupal Dev, P.~Millington, A.~Pilaftsis and D.~Teresi,
  %``Flavour Covariant Transport Equations: an Application to Resonant Leptogenesis,''
  Nucl.\ Phys.\ B {\bf 886} (2014) 569
  [arXiv:1404.1003 [hep-ph]];
  P.~S.~Bhupal Dev, P.~Millington, A.~Pilaftsis and D.~Teresi,
  %``Kadanoff–Baym approach to flavour mixing and oscillations in resonant leptogenesis,''
  Nucl.\ Phys.\ B {\bf 891} (2015) 128
  [arXiv:1410.6434 [hep-ph]];  

\bibitem{D'Onofrio:2014kta}
  M.~D'Onofrio, K.~Rummukainen and A.~Tranberg,
  %``Sphaleron Rate in the Minimal Standard Model,''
  Phys.\ Rev.\ Lett.\  {\bf 113} (2014) no.14,  141602
  [arXiv:1404.3565 [hep-ph]].

\bibitem{Carrington:1991hz}
  M.~E.~Carrington, 
  %``The Effective potential at finite temperature in the Standard Model,''
  Phys.\ Rev.\ D {\bf 45} (1992) 2933;
%\bibitem{Kajantie:1995kf}
  K.~Kajantie, M.~Laine, K.~Rummukainen and M.~E.~Shaposhnikov,
  %``The Electroweak phase transition: A Nonperturbative analysis,''
  Nucl.\ Phys.\ B {\bf 466} (1996) 189
  [hep-lat/9510020].

\bibitem{Giudice:2003jh}
  G.~F.~Giudice, A.~Notari, M.~Raidal, A.~Riotto and A.~Strumia,
  %``Towards a complete theory of thermal leptogenesis in the SM and MSSM,''
  Nucl.\ Phys.\ B {\bf 685} (2004) 89
  [hep-ph/0310123].
  
\bibitem{Espinosa:1993bs}
  J.~R.~Espinosa and M.~Quiros,
  %``The Electroweak phase transition with a singlet,''
  Phys.\ Lett.\ B {\bf 305} (1993) 98
  [hep-ph/9301285].    
  

\bibitem{future} In preparation.  

  
\bibitem{IR}
  D.~Besak and D.~Bodeker,
  %``Thermal production of ultrarelativistic right-handed neutrinos: Complete leading-order results,''
  JCAP {\bf 1203} (2012) 029 [arXiv:1202.1288 [hep-ph]];  
%\bibitem{Ghiglieri:2016xye} 
  J.~Ghiglieri and M.~Laine,
  %``Neutrino dynamics below the electroweak crossover,''
  arXiv:1605.07720 [hep-ph].
  


 
 
\bibitem{Frossard:2012pc}
  T.~Frossard, M.~Garny, A.~Hohenegger, A.~Kartavtsev and D.~Mitrouskas,
  %``Systematic approach to thermal leptogenesis,''
  Phys.\ Rev.\ D {\bf 87} (2013) no.8,  085009
  [arXiv:1211.2140 [hep-ph]].  
  
  
\bibitem{Kartavtsev:2015vto}
  A.~Kartavtsev, P.~Millington and H.~Vogel,
  %``Lepton asymmetry from mixing and oscillations,''
  JHEP {\bf 1606} (2016) 066
  [arXiv:1601.03086 [hep-ph]].
 
%\bibitem{Dev:2014wsa}
%  P.~S.~Bhupal Dev, P.~Millington, A.~Pilaftsis and D.~Teresi,
%  %``Kadanoff–Baym approach to flavour mixing and oscillations in resonant leptogenesis,''
%  Nucl.\ Phys.\ B {\bf 891} (2015) 128
%  [arXiv:1410.6434 [hep-ph]];  
%\bibitem{Kartavtsev:2015vto}  
%  A.~Kartavtsev, P.~Millington and H.~Vogel,
%  %``Lepton asymmetry from mixing and oscillations,''
%  arXiv:1601.03086 [hep-ph].
%  %%CITATION = ARXIV:1601.03086;%%
%  %1 citations counted in INSPIRE as of 07 Apr 2016

\bibitem{Hohenegger:2014cpa}
  A.~Hohenegger and A.~Kartavtsev,
  %``Leptogenesis in crossing and runaway regimes,''
  JHEP {\bf 1407} (2014) 130
  [arXiv:1404.5309 [hep-ph]].
  

\bibitem{Alekhin:2015byh}
  S.~Alekhin {\it et al.},
  %``A facility to Search for Hidden Particles at the CERN SPS: the SHiP physics case,''
  arXiv:1504.04855 [hep-ph].
  
\bibitem{FCC}
  A.~Blondel {\it et al.} [FCC-ee study Team Collaboration],
  %``Search for Heavy Right Handed Neutrinos at the FCC-ee,''
  arXiv:1411.5230 [hep-ex]; 
%\bibitem{Antusch:2016vyf}  
  S.~Antusch, E.~Cazzato and O.~Fischer,
  %``Displaced vertex searches for sterile neutrinos at future lepton colliders,''
  arXiv:1604.02420 [hep-ph].


  
  
\bibitem{Drewes:2012ma}
  M.~Drewes and B.~Garbrecht,
  %``Leptogenesis from a GeV Seesaw without Mass Degeneracy,''
  JHEP {\bf 1303} (2013) 096
  [arXiv:1206.5537 [hep-ph]].

%\cite{Akhmedov:1998qx}
\bibitem{Akhmedov:1998qx}
  E.~K.~Akhmedov, V.~A.~Rubakov and A.~Y.~Smirnov,
  %``Baryogenesis via neutrino oscillations,''
  Phys.\ Rev.\ Lett.\  {\bf 81} (1998) 1359
  [hep-ph/9803255].
  %%CITATION = doi:10.1103/PhysRevLett.81.1359;%%  
  
  
\bibitem{Asaka:2005pn}
  T.~Asaka and M.~Shaposhnikov,
  %``The nuMSM, dark matter and baryon asymmetry of the universe,''
  Phys.\ Lett.\ B {\bf 620} (2005) 17
  [hep-ph/0505013].
 
  
\bibitem{Canetti:2012kh}
  L.~Canetti, M.~Drewes, T.~Frossard and M.~Shaposhnikov,
  %``Dark Matter, Baryogenesis and Neutrino Oscillations from Right Handed Neutrinos,''
  Phys.\ Rev.\ D {\bf 87} (2013) 093006
  [arXiv:1208.4607 [hep-ph]].

  

  
  

  

  
\end{thebibliography}
\end{document}